\documentclass[draftcls,9pt,conference]{IEEEtran}

\usepackage[dvips]{color}
\usepackage{graphicx}
\usepackage[T1]{fontenc}
\usepackage[francais]{babel}

\usepackage{amsmath,epsfig}
\usepackage{epsfig}
\usepackage{amsmath,amstext,amsfonts,amssymb}
\usepackage{epsfig,float}
\usepackage{graphicx}
\usepackage{subfigure}
\usepackage{vmargin}
\setmarginsrb{2.cm}{2.35cm}{2.0cm}{2.0cm}{0cm}{0cm}{0cm}{0.5cm}

\newcommand{\ul}{\underline}

\newcommand{\mc}{\mathcal}

\newcommand{\ds}{\displaystyle}

\begin{document}

\title{Combining Coded Signals with Arbitrary Modulations in Orthogonal Relay Channels}

\author{
\authorblockN{Brice Djeumou}
\authorblockA{Lab. des Signaux et Syst\`{e}mes \\
CNRS - Sup\'{e}lec - Univ Paris Sud \\
91191 Gif-sur-Yvette, France \\
djeumou@lss.supelec.fr}
 \and
\authorblockN{Samson Lasaulce}
\authorblockA{Lab. des Signaux et Syst\`{e}mes \\
CNRS - Sup\'{e}lec  - Univ Paris Sud \\
91191 Gif-sur-Yvette, France \\
lasaulce@lss.supelec.fr}\\
\and
\authorblockN{Antoine O. Berthet}
\authorblockA{Dpt. T\'{e}l\'{e}coms\\
Sup\'{e}lec\\
91191 Gif-sur-Yvette, France \\
antoine.berthet@supelec.fr}}

\maketitle

\begin{abstract}
We consider a relay channel for which the following assumptions are made: 1. The source-destination and relay-destination
channels are orthogonal (frequency division relay channel); 2. The relay implements the decode-and-forward protocol; 3. The
source and relay implement the same channel encoder namely a convolutional encoder; 4. They can use arbitrary and possibly
different modulations. In this framework, we derive the best combiner in the sense of the maximum likelihood (ML) at the
destination and the branch metrics of the trellis associated with its channel decoder for the ML combiner and also
 for the maximum ratio combiner (MRC),
Cooperative-MRC (C-MRC) and the minimum mean square error (MMSE) combiner.
\end{abstract}

%%%%%%%%%%%%%%%%%%%%%%%%%%%%%%%%%%%%%%%%%%%%%%%%%%%%%%%%%%%%%%%%%%%%%%%%%%%%%%%%%%%%%%%%%%%%%%%%%%%%%%%%%%%%%%%%%%%%%%%%%%%%%%%%%%%%%%%%%%%%%   Section 1 : Intro    %%%%%%%%%%%%%%%%%%%%%%%%%%%%%%%%%%%%%%%%%%%%%
%%%%%%%%%%%%%%%%%%%%%%%%%%%%%%%%%%%%%%%%%%%%%%%%%%%%%%%%%%%%%%%%%%%%%%%%%%%%%%%%%%%%%%%%%%%%%%%%%%%%%%%%%
%----------------------------------------------------------------------
\section{Motivations and technical background}
\label{intro}

We consider orthogonal relay channels for which orthogonality is implemented in frequency \cite{elgamal-it-2006}. Since the
source-destination channel is assumed to be orthogonal to the relay-destination channel the destination receives two distinct
signals. In order to maintain the receiver complexity at a low level, the destination is imposed to combine the received
signals before applying channel decoding. The relay is assumed to implement the decode-and-forward (DF) protocol. We have at
least two motivations for this choice. First, in contrast with the well-known amplify-and-forward (AF) protocol, it can be
implemented in a digital relay transceiver. More importantly, whereas the AF protocol imposes the source-relay channel to have
the same bandwidth as the relay-destination channel, the DF protocol offers some degrees of freedom in this respect. This is a
critical point when the cooperative network has to be designed from the association of two existing networks. For instance, if
one wants to increase the performance of a Digital Video Broadcasting (DVB) receiver or reach some uncovered indoor areas, a
possible solution is to use cell phones, say Universal Mobile Telecommunications System (UMTS) cell phones as relaying nodes.
The problem is that DVB signals use a 20 MHz bandwidth (source-relay channel) while UMTS signals have only a bandwidth of 5 MHz
(relay-destination channel). The AF protocol cannot be used here. But the DF protocol can be used, for instance, by adapting
the modulation of the cooperative signal to the available bandwidth. In this case, the destination has to combine two signals
with different modulations.

In this context, one of the issues that needs to be addressed is the
design of the combiner. A conventional MRC cannot be used for
combining signals with different modulations (except for special
cases of modulations). Even if the modulations at the source and
relay are identical, the MRC can severely degrade the receiver
performance because it does not compensate for the decoding noise
introduced by the relay
\cite{laneman-wcnc-2000}--\cite{wang-com-2007}. This is why the
authors of \cite{laneman-wcnc-2000}\cite{djeumou-isspit-2006}
proposed a maximum likelihood detector (MLD) for combining two
BPSK-modulated signals coming from the source and relay. The authors
of \cite{wang-com-2007} proposed an improved MRC called C-MRC which
aims at maximizing receive diversity. The authors
 of \cite{sendonaris-com-2003} proposed a linear combiner for which the weights are tuned
 to minimize the raw bit error rate (BER). The main issue
is that one is not always able to explicit the raw BER as a function
of the combiner weights whereas the likelihood
 calculation is more systematic. Additionally, when some a
priori knowledge is available, the ML metric can be used to calculate an a posteriori probability (APP). In the context of
orthogonal $N-$relay channels the authors of \cite{djeumou-isspit-2006} derived two new combiners: the best MRC in the sense of
the equivalent signal-to-noise ratio and MMSE combiner. They also assessed the BER performance of the latter and MLD in the
uncoded case.

Compared to the aforementioned works, this paper also aims at designing a good combiner at the destination but it differs from
them on two essential points: 1. The interaction between the combiner and channel decoder is exploited in the sense that we
want to express the branch metrics of the trellis associated with channel decoding for the MRC, MMSE combiner, C-MRC and
especially for the ML combiner; 2. When the ML combiner is assumed, the source and relay can use arbitrary modulations (not
necessarily BPSK modulations as in \cite{laneman-wcnc-2000}\cite{sendonaris-com-2003}\cite{chen-wireless-2006}) and, more
importantly, these can be different.

%----------------------------------------------------------------------
\section{Signal model}
\label{sm-cs}

At the source the $L$-information bit sequence $\ul{m}$ is encoded into a sequence of bits $\ul{b}$ and modulated into the
transmitted signal $\ul{x} = \left(x(1),...,x(T) \right)$ where, $\forall t \in \{1,...,T \}, \ x(t) \in \mc{X}$, $\mc{X}$ is a
finite alphabet corresponding to the modulation constellation used by the source and $ \mathbb{E}\left|x(t) \right|^2 \leq
P_0$. At the relay the message is decoded, re-encoded with the same encoder as the source and modulated into the transmitted
signal $\ul{x}_1 = \left(x_1(1),...,x_1(T_1) \right)$ where, $\forall  t_1 \in \{1,...,T_1 \}, \ x_1(t_1) \in \mc{X}_1$,
 $\mc{X}_1$ is a finite
alphabet corresponding to the modulation constellation used by the relay and
 $ \mathbb{E}\left|x_1(t_1) \right|^2 \leq P_1$. We denote by $s$ (resp. $r$) the number of coded bits conveyed by
one source (resp. relay) symbol. By definition: $s = \log_2|\mc{X}|$ and $r = \log_2|\mc{X}_1|$. More specifically, the
information bit sequence is assumed to be encoded by a $\frac{1}{q}$-rate convolutional encoder ($q \in \mathbb{N}^*$). As the
sequence $\ul{x}$ comprises $T$ symbols we have that $q(k+ \nu) = sT$ where $\nu$ is the channel encoder memory. Assuming time
selective but frequency non-selective channels, the baseband signals received by the destination from the source and relay
respectively write $y_0(t) = h_0 x(t) + z_0(t)$ and
 $y_1(t_1) = h_1 x_1(t_1) + z_1(t_1)$ where
$z_0$ and $z_1$ are zero-mean circularly symmetric complex Gaussian noises with
 variances $\sigma_{0}^2$ and $\sigma_{1}^2$
respectively. The complex coefficients $h_{0}$ and $h_{1}$ represent the gains of the source-destination and source-relay
fading channels. For insuring coherent decoding, these two gains are assumed to be known to the receiver and relay
respectively. We define $\gamma_0 = \mathbb{E}\left|h_0\right|^2\frac{ P}{\sigma_0^2}$, $\gamma_1 =
\mathbb{E}\left|h_1\right|^2\frac{ P}{\sigma_1^2}$, $\gamma_1^{'} = \mathbb{E}\left|h_1^{'}\right|^2\frac{ P}{\sigma_0^2}$
 and $\rho_1 = \frac{\mathbb{E}\left|X_1X^{*}\right|}{P}$ where $h_1^{'}$ is the gain of the source-relay fading channel. Note that, in order to ensure the conservation of the coded bit
rate between the input and output of the relay, $s$ and $r$ have to be linked by the following compatibility relation: $sT = r
T_1$. In the sequel we will use the quantity $k = \textrm{lcm}(s,r)$ (where $\textrm{lcm}$ is the least common multiple
function). For simplicity, we assume that the source and relay use the same channel coder. Therefore the relay has to use a
modulation that is compatible with the source's one. We will also assume that the number of times per second the channel can be
used is directly proportional to the available bandwidth. For example, if the source uses a BPSK modulation and the cooperation
channel has a bandwidth equal to half the downlink bandwidth, the relay can use a QPSK modulation.

\section{A new trellis branch metric}

\subsection{When the source and relay use arbitrary and different modulations}\label{mld}

In this case, the linear combiners derived by \cite{sendonaris-com-2003}\cite{djeumou-isspit-2006}\cite{wang-com-2007} cannot
be used in general. However, provided that the above compatibility condition is met, the ML combiner can be derived as we show
now. Let us denote by $\ul{y}_{0}$ and $\ul{y}_{1}$ the sequences of noisy symbols
 received by the destination from the source
and relay respectively.  The discrete optimization problem the ML combiner solves is as follows:
\begin{eqnarray*}
\widehat{\ul{m}}= \arg \ds{ \max_{\ul{m} \in \Bbb{F}_{2}^{L}}} \ p_{ML} = \arg \ds{ \max_{\ul{m}\in \Bbb{F}_{2}^{L}}} p\left(
\ul{y}_{0},\ul{y}_{1}|\ul{x}\right). %\arg \ds{ \max_{\ul{m}\in \Bbb{F}_{2}^{L}}} p\left(\ul{y}_{0},\ul{y}_{1}|\ul{m}\right) =
\label{eq:arg-max}
\end{eqnarray*}
As the reception
 noises are assumed to be independent,  $ p_{ML} =
p\left(\ul{y}_{0}|\ul{x}\right)p\left(\ul{y}_{1}|\ul{x}\right)$. The first term easily writes as
\begin{eqnarray*}
p\left(\ul{y}_{0}|\ul{x}\right)  = \ds{\prod_{t=1}^{T} \frac{1}{\pi\sigma_{0}^2}} \
 \exp\left(-\frac{\left|{y}_{0}(t)- h_0 x(t)\right|^2}{\sigma_{0}^2}\right). %= \ds\prod_{t=1}^{T} p\left({y}_{0}(t)|{x}(t)\right)
 \label{eq:ml-sd}
\end{eqnarray*}
 In order to express the second term, we introduce a sequence of $T_1$ discrete symbols
denoted by $\ul{e}_1$ which models the residual noise at the relay after the decoding--re-encoding process. This noise is
therefore modeled by a multiplicative error term which is not independent of the symbols transmitted by the relay.
Additionally, the statistics of this noise are assumed to be known by the destination. For this, one can establish once and for
all a lookup table between the source-relay SNR and the bit error rate after re-encoding at the relay. The cooperation signal
writes then
 as $y_1(t_1) = h_1 x_1(t_1) + z_1(t_1)$ where $x_1(t_1) =
\ul{e}_1(t_1) \tilde{x}_1(t_1)$ and $\tilde{x}_1(t_1)$ is the symbol the relay would generate if there were no decoding error
at the relay. For example, when the relay uses a QPSK modulation, $e_1 \in \{1, e^{j\frac{\pi}{2}}, e^{j\pi}, e^{j\frac{3
\pi}{2}}\}$. Therefore we have that $p\left(\ul{y}_{1}|\ul{x}\right) = p\left(\ul{y}_{1}|\tilde{\ul{x}}_1\right)
=\ds{\sum_{\ul{e}_1}}p\left(\ul{y}_{1},\ul{e}_1|\tilde{\ul{x}}_1\right) =
\ds{\sum_{\ul{e}_1}}p\left(\ul{e}_1|\tilde{\ul{x}}_1\right) p\left(\ul{y}_{1}|\tilde{\ul{x}}_1,\ul{e}_1\right)$. At this point,
we need to make an additional assumption in order to easily derive the path metric of the ML decoder. From now one, we assume
that the discrete symbols of the sequence $\ul{e}_1$ are conditionally independent. This assumption is very realistic, for
example, if the source and relay implement a bit interleaved coded modulation (BICM) or a trellis coded modulation (TCM). In
the case of the BICM, the channel coder, which generates coded bits, and the modulator are separated by an interleaver. The
presence of this interleaver precisely makes the proposed assumption reasonable. Under the aforementioned assumption one can
expand $p\left(\ul{y}_{1}|\ul{x}\right)$ as
\begin{eqnarray*}
\begin{array}{lll}
p\left(\ul{y}_{1}|\ul{x}\right) = \ds{\sum_{\ul{e}_1}}\prod_{t_1=1}^{T_1} p\left({e}_1(t_1)|\tilde{{x}}_1(t_1)\right)
p\left({y}_{1}(t_1)|\tilde{{x}}_1(t_1),{e}_1(t_1)\right)&&\\
 \ \ \ \ \ \ \ \ \ \ \ \ = \ds{\sum_{\ul{e}_1}\prod_{t_1=1}^{T_1}} p\left({e}_1(t_1)|\tilde{{x}}_1(t_1)\right) \times&& \\
 \ \ \ \ \ \ \ \ \ \ \ \ \ \ \ \ \ \ \ \ \  \ \ \ \frac{1}{\pi\sigma_{1}^2} \exp\left(-\frac{|{y}_{1}(t_1)- h_1
{e}_1(t_1)\tilde{x}_1(t_1)|^2}{\sigma_{1}^2}\right).&&
\end{array} \nonumber
\label{eq-ml-srd}
\end{eqnarray*}
The main consequence of this assumption is a significant reduction of the decoder complexity. If the assumption is not valid,
the proposed derivation can always be used but the performance gain obtained can be marginal since the errors produced by will
not be spread over the data block but rather occurs in a sporadic manner along the block.

In order to express the path metric of a given path in the trellis associated with channel decoding, we need now to link the
likelihood expressed above and the likelihood associated with a given bit $b_j$, where $j \in \{1,..., k\}$. The reason why we
consider sub-blocks of $k$ bits is that, in order to meet the rate compatibility condition, the ML combiner combines the $k_s =
\frac{k}{s}$ symbols received from the source with the $k_r = \frac{k}{r}$ symbols received from the relay. Now, $\forall (i,j)
\in \{0,1\}\times\{1,...,k\}$, let us define the sets $\mc{B}_{i}^{(k)}(j) = \{\ul{b}^{k} = (b_1,...,b_{k}) \in \{0,1\}^{k}, \
b_j = i\}$, a set of sub-blocks of $rs$ consecutive bits, $\mc{X}_{i}^{(k_s)}(j) = \{\ul{x}^{k_s} \in \mc{X}^{k_s}, \ b_j =
i\}$ and $\mc{X}_{1,i}^{(k_r)}(j) = \{\ul{\tilde{x}}_{1}^{k_r} \in \mc{X}_1^{k_r}, \ b_j = i\}$, their equivalents in the
source (resp. relay) modulation space. With these notations the bit likelihood can be expressed as follows
\begin{eqnarray*}
\begin{array}{l}
\lambda \left(b_{j}=i\right) \ = \ \ds{\ln\big[\ds{\sum_{\ul{b}_1^{k} \in \mc{B}_{i}^{(k)}(j)}} p\left(\ul{y}_{0,1}^{k_s}|
\ul{x}_{1}^{k_s} \right) p\left({y}_{1,1}^{k_r}|\ul{\tilde{x}}_{1,1}^{k_r}\right)\big]} \\
 =  \ds{ \ln\big[\sum_{\ul{b}_1^{k} \in \mc{B}_{i}^{(k)}(j)} \prod_{t=1}^{k_s} p\left({y}_{0}(t)|{x}(t)\right)
\prod_{t_1=1}^{k_r} p\left({y}_{1}(t_1)|\tilde{x}_1(t_1)\right)\big]},
\end{array}
\label{eq:llr1}
\end{eqnarray*}
where we used the notation $\ul{v}_1^n = (v(1),...,v(n))$. When a BICM is used, the obtained log-likelihood sequence is then
de-interleaved and given to a Viterbi decoder.

%-------------------------------------------------------------------------------------------------
%-------------------------------------------------------------------------------------------------
\subsection{When the source and relay use arbitrary and identical modulations}
\label{metric-others}

The derivation of the coded-bit likelihood in the case where the modulations used by the source and relay are the same is ready
since it is special case of derivation conducted previously with $k=s=r$. In this case, both ML and linear combiners can be
used since the combination can be performed symbol-by-symbol. The log-likelihood becomes $\lambda\left(b_{j}=i\right) =
\ln\left[\ds{\sum_{\ul{b}_1^{s} \in \mc{B}_{i}^{(s)}(j)}} p\left({y}_{0}(t)|{x}(t)\right)
p\left({y}_{1}(t)|{x}(t)\right)\right]$, where $1 \leq j \leq s$. If we further assume that the modulations used are BPSK
modulations, the likelihood on the received sequences take a more explicit form. Indeed, it can be checked that
\begin{eqnarray*}
\begin{array}{l}
\ln\left[p\left(\ul{y}_{0},\ul{y}_{1}|\ul{m}\right)\right] =  \\
 \ds{-kq\ln\left(2\pi\right) - T\ln\left(\sigma_{0}^2\sigma_{1}^2\right) -
\sum_{t=1}^{T}\left[\frac{\left|{y}_{0}(t)- h_0 x(t)\right|^2}{\sigma_{0}^2} \right. } \\
\ds{ \left. - \ln\left[\sum_{e=-1,+1}
\mathrm{Pr}\left[\epsilon(t)=e\right]
\exp\left(-\frac{\left|{y}_{1}(t)-
h_1ex(t)\right|^2}{\sigma_{1}^2}\right)\right] \right]}
\end{array}
    \label{eq:ln-mlc}
\end{eqnarray*}
where $\mathrm{Pr}[\epsilon=-1]$ represents the residual bit error rate (after the decoding--re-encoding procedure inherent to
DF protocol). Denote by $\pi$ the interleaver function such that $t = \pi\left(t_0\right)$ and $t_0 = \pi^{-1}\left(t\right)$.
Finally, the path metric is merely given by
\begin{eqnarray*}
\begin{array}{l}
\mu^{\bold{x}}  = \ds{\sum_{t_0=1}^{T} \left[\frac{\left|{y}_{0}(t_0)-h_0x(t_0)\right|^2}{\sigma_{0}^2} -\right. }\\
\ds{\left. \ln\left[\sum_{e=-1,+1}  P_r\left(\epsilon(t_0)=e\right)
\exp\left(\frac{\left|{y}_{1}(t_0)-h_1ex(t_0)\right|^2}{\sigma_{1}^2}\right)\right]
\right]}.
\end{array}
    \label{eq:c-metrique}
\end{eqnarray*}
So the combining and channel decoding are performed jointly by modifying the branch metrics as indicated
above.\\
When using a linear combiner, one has to compute the APP from the equivalent signal at the combiner output. This computation
requires the equivalent channel gain and noise. We provide them for each linear combiner considered here. For a given combiner,
denote its optimal vector of weights by $\ul{w} = (w_0,w_1)$ and rewrite the signal at the combiner output as $y =
\ds{\sum_{i=0}^{1} \ w_i y_i = h_{eq} x+ z_{eq}}$, where $h_{eq}$ and $z_{eq} \sim \mc{N}(0, \sigma_{eq}^2)$ are the equivalent
channel gain and noise respectively. The bit
 log-likelihood can then be easily expressed as $\lambda\left(b_{j}=i\right)   = \ln\big[\ds{\sum_{\ul{x} \in
\mc{X}^{(s)}_{i}(j)} } p\left(y \ | \ h_{eq},\ul{x} \right)\big]$. Table \ref{tab:eq-gain-noise} summarizes the values of these
quantities with the notations $a_0 = |h_{0}|^2$ and $a_1 = |h_{1}|^2$.
\begin{table}[h!]
    %\centering
        \begin{tabular}{|l|c|c|} \hline
                                                        & $h_{eq}$                          &  $\sigma_{eq}^2$   \\\hline
         MRC                         & $\frac{a_0}{\sigma_{0}^2} + \frac{a_1}{\sigma_1^2}$
             &   $\frac{a_0}{\sigma_0^2} + \frac{a_1}{\sigma_1^2}$  \\\hline
         MMSE     & $\frac{a_0}{\sigma_0^2} + \frac{a_1 \rho_{1}^2}{\sigma_1^2
         + a_1 P_{0}\left(\alpha_{1}^2-\rho^2_{1}\right)}$
               &   $\frac{a_0}{\sigma_0^2} + \frac{a_1 \rho_{1}^2}{\sigma_1^2 +
               a_1 P_{0}\left(\alpha_{1}^2-\rho^2_{1}\right)}$
               \\\hline
         C-MRC                       & $a_0 + \frac{\min\{\gamma_{1}',\gamma_{1}\}}{\gamma_{1}} a_1$
           &   $a_0 \sigma_0^2 + \frac{\min\{\gamma_{1}',\gamma_{1}\}}{\gamma_{1}} a_1 \sigma_1^2$
            \\\hline
        \end{tabular} \\
    \caption{Equivalent channel parameters for the linear combiners} \vspace{-1cm}
    \label{tab:eq-gain-noise}
\end{table}

%%%%%%%%%%%%%%%%%%%%%%%%%%%%%%%%%      Section 3 : Simulation results   %%%%%%%%%%%%%%%%%%%%%%%%%%%%%%%%
%%%%%%%%%%%%%%%%%%%%%%%%%%%%%%%%%%%%%%%%%%%%%%%%%%%%%%%%%%%%%%%%%%%%%%%%%%%%%%%%%%%%%%%%%%%%%%%%%%%%%%%%
\section{Simulation example}
\label{sim} 
\begin{figure}[h!]
  \begin{center}
%%%
    \includegraphics[width=1.1\columnwidth]
    {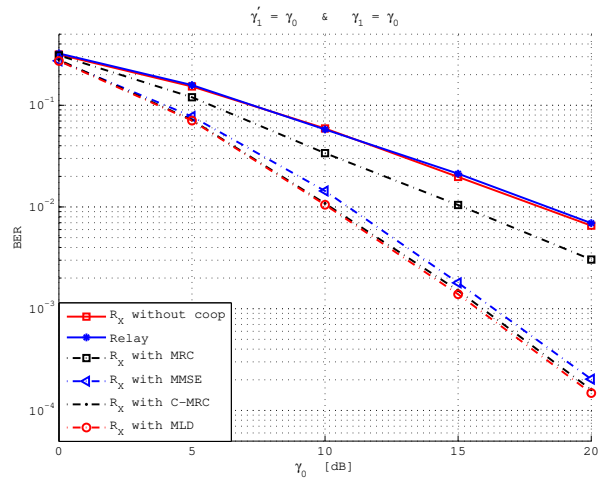}
  \end{center}
  \vspace{-1cm}
  \caption{BER at the destination with $\gamma_{1}' = \gamma_{0}$, $\gamma_{1}=\gamma_{0}$}
  \label{fig:1}
\end{figure}

\begin{figure}[h!]
  \begin{center}
%%%
    \includegraphics[width=1.1\columnwidth]
    {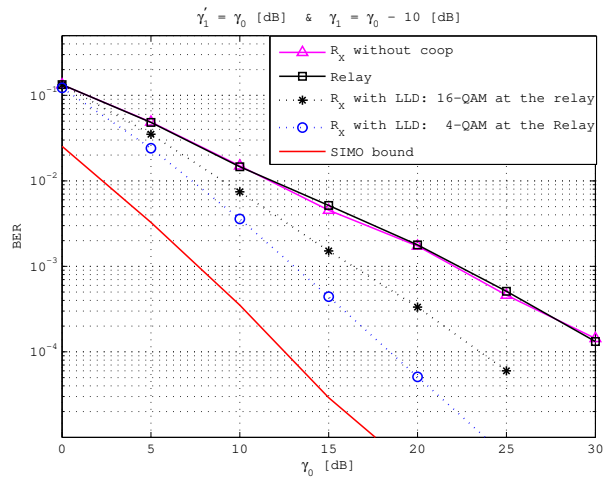}
  \end{center}
  \vspace{-1cm}
  \caption{BER at the destination $\gamma_{1}' = \gamma_{0}$, $\gamma_{1}$ = $\gamma_{0} - 10$ \ $\mathrm{dB}$}
  \label{fig:2}
\end{figure}

For the 2 figures provided, we assume that the source and the relay implement a $\frac{1}{2}-$rate convolutional
encoder ($4-$state encoder with a free distance equal to $5$). Frequency non-selective Rayleigh block fading channels are
assumed and the data block length is chosen to be $1024$. First, we compare the combiners between themselves when both the
relay and source use a $4-$QAM modulation. Fig. \ref{fig:1} represents the BER at the decoder output as a function of
$\gamma_0=\gamma_1$. There are 6 curves. From the top to the bottom, they respectively represent the performance with no relay,
with the relay associated with the conventional MRC, MMSE, C-MRC and ML combiners. When implementing the conventional MRC, the
receiver does not significantly improve its performance w.r.t. to the non cooperative case whereas the other combiners can
provide more than a $8$ dB gain and perform quite similarly. Then (see Fig. \ref{fig:2}), we evaluate the performance gain
brought by the MLD when the source and relay have to use different modulations: the source implements a BPSK while the relay
implements either a $4-$QAM or a $16-$QAM. The second scenario would correspond to a case where the source-destination channel
bandwidth is 4 times larger than the source-destination channel bandwidth (\emph{e.g.} 20 MHz vs 5 MHz). We see that the MLD
not only makes cooperation possible but also allows the destination to extract a significant performance gain from it. To have
an additional reference, we also represented the performance of the equivalent virtual $1 \times 2 $ MIMO system, which is
obtained for $\gamma_1 = + \infty$.

%%%%%%%%%%%%%%%%%%%%%%%%%%%%%%%%%%%%%%%%%%%%%%%%%%%%%%%%%%%%%%%%%
%%%%%%%%%%%%%%%%%%%%%%%%%%%%%%%%%%%%%%%%%%%%%%%%%%%%%%%%%%%%%%%%%%%%%%%%%%%%%%%%%%%%%%%%%%%%%%%%%%%%%%%%
\section{Concluding remarks}
\label{sec:concl} The results provided in this letter and many other simulations performed in the \emph{coded} case led us to
the following conclusion: if the source and relay can use the same modulation, the C-MRC generally offers the best
performance-complexity trade-off. On the other hand, if the modulations are \emph{different}, as it would be generally the case
when two \emph{existing} communications systems are associated to cooperate, linear combiners and thus the C-MRC cannot be used
in general and the ML combiner is the only implementable combiner.

%\bibliography{pimrc}
{}

\end{document}